\pdfoutput=1
\documentclass[sigconf]{acmart}
\usepackage{float}
\usepackage{multicol}
\usepackage{balance}

\usepackage{epsfig}
\usepackage{booktabs} % For formal tables
\usepackage{adjustbox}
\usepackage{enumitem}
\usepackage{eurosym}

\usepackage{graphicx}

\usepackage[utf8]{inputenc}
\usepackage[T1]{fontenc}

\begin{document}
    \title{Does Facebook Use Sensitive Data for Advertising Purposes?}
    \subtitle{Worldwide Analysis and GDPR Impact}

%\author{Ángel Cuevas,
%      José González Cabañas,
%      Aritz Arrate, and
%      Rubén Cuevas}
      %Department of Telematic Engineering, Universidad Carlos III de %Madrid}  
%\affiliation{%
%  \institution{Department of Telematic Engineering, Universidad Carlos III de Madrid}
%  \streetaddress{Av. de la Universidad, 30}
%  \city{Leganés}
%  \state{Madrid}
%  \country{Spain}
%  \postcode{28911}
%}
%\email{{acrumin, jgcabana}@it.uc3m.es, aarrate@pa.uc3m.es, rcuevas @it.uc3m.es}

% The paper headers
% \markboth{IEEE TRANSACTIONS ON PRIVACY
% }{A. Cuevas \MakeLowercase{\textit{et al.}}: World Extension Sensitive Interests}

\author{Ángel Cuevas}
\affiliation{%
  \institution{Universidad Carlos III de Madrid}}
\email{acrumin@it.uc3m.es}

\author{José González Cabañas}
\affiliation{%
  \institution{Universidad Carlos III de Madrid}}
\email{jgcabana@it.uc3m.es}

\author{Aritz Arrate}
\affiliation{%
  \institution{Universidad Carlos III de Madrid}}
\email{aarrate@pa.uc3m.es}

\author{Rubén Cuevas}
\affiliation{%
  \institution{Universidad Carlos III de Madrid}}
\email{rcuevas@it.uc3m.es}

% === ABSTRACT ====================================================================
% =================================================================================
\begin{abstract}
%\boldmath
%\vspace{-0.25em}
The recent European General Data Protection Regulation (GDPR) and other data protection regulations restrict the processing of some categories of personal data (health, political orientation, sexual preferences, religious beliefs, ethnic origin, etc.) due to the privacy risks associated to such information. The GDPR refers to these categories as sensitive personal data. This paper quantifies the portion of Facebook (FB) users, across 197 countries, who are labeled with advertising interests linked to potentially sensitive personal data. Our study reveals that Facebook labels 67\% of users with potential sensitive interests. This corresponds to 22\% of the population in the referred 197 countries. Moreover, our work shows that the GDPR enforcement had a negligible impact in this context since the portion of FB users labeled with sensitive interests in the European Union remains almost the same 5 months before and 9 months after the GDPR was enacted. The paper also illustrates potential risks associated to the use of sensitive interests. For instance, we quantify the portion of FB users labelled with the interest \textit{"Homosexuality"} in countries where being gay may be punished with the death penalty. The last contribution is the implementation of a web browser extension that allows FB users removing in a simple way the potentially sensitive interests FB has assigned them.
\end{abstract}

\settopmatter{printacmref=false}
\renewcommand\footnotetextcopyrightpermission[1]{} % removes footnote with conference information in first column
\pagestyle{plain} % removes running headers

\maketitle

% ====================================================================
% ====================================================================
% ====================================================================
 
\section{Introduction}

%The irruption of very popular online services whose business model builds up on the commercial exploitation of personal information through tailored advertising and personalized recommendation of products, services or content has raised a very intense debate around questions like where are the ethical and legal boundaries in the management of personal information. In 2012, The World Economic Forum \cite{WorldEconomicForum} concluded that the lack of the increasing concern of citizens regarding privacy and data protection was a serious risk for the sustainable economic growth of online services. 

Worldwide citizens have demonstrated serious concerns regarding the management of personal information by online services. For instance, the 2015 Eurobarometer about data protection \cite{Eurobarometer} reveals that: 63\% of EU citizens do not trust online businesses, more than half do not like providing personal information in return for free services, and 53\% do not like that Internet companies use their personal information in tailored advertising. Similarly, a recent survey carried out among US users \cite{janrain} reveals that 53\% of respondents were against receiving tailored ads from the information websites and apps learn about them, 42\% do not think websites care about using users data in a secure and responsible way at all, and 73\% considers web sites know too much about users. A survey conducted by Internet Society (ISOC) in the Asia-Pacific region in 2016 \cite{internetsociety} disclosed that 59\% of the respondent did not feel their privacy is sufficiently protected when using the Internet and 45\% considered urgent to get the attention of policymakers in their country on data protection matters.  

Policymakers have reacted to this situation by passing or proposing new regulations in the area of privacy and/or data protection. For instance, in May 2018 the EU enforced the General Data Protection Regulation (GDPR) \cite{GDPR} across all 28 member states. Similarly, in June 2018 California passed the California Consumer Privacy Act \cite{california}, claimed to be the nation's toughest data privacy law. In countries like Argentina or Chile governments proposed in 2017 new bills updating their existing data protection regulation \cite{pwc}. For the purpose of this paper we will take as reference the GDPR since it is the one affecting more countries, citizens and companies.

The GDPR (but also most data protection regulations) defines some categories of personal data as sensitive and prohibits processing them with limited exceptions (\textit{e.g.,} the user provides explicit consent to process that sensitive data for a specific purpose). In particular, the GDPR defines as sensitive personal data:  \textit{``data revealing racial or ethnic origin, political opinions, religious or philosophical beliefs, or trade union membership, and the processing of genetic data, bio-metric data for the purpose of uniquely identifying a natural person, data concerning health or data concerning a natural person's sex life or sexual orientation''}.  

Due to the legal, ethical and privacy implications of processing sensitive personal data, it is important to verify whether online services may be commercially exploiting such sensitive information. If that is the case, it is also essential to measure the portion of users/citizens who may be affected by the exploitation of their sensitive personal data. In this paper, we address these crucial questions focusing on \emph{online advertising}, which represents the most important source of revenue for most online services. In particular, we consider the case of Facebook (FB), whose online advertising platform is second only to Google in terms of revenue \cite{FB2nd}. 

\begin{figure}[b]
\centering
\includegraphics[width=.9\columnwidth]{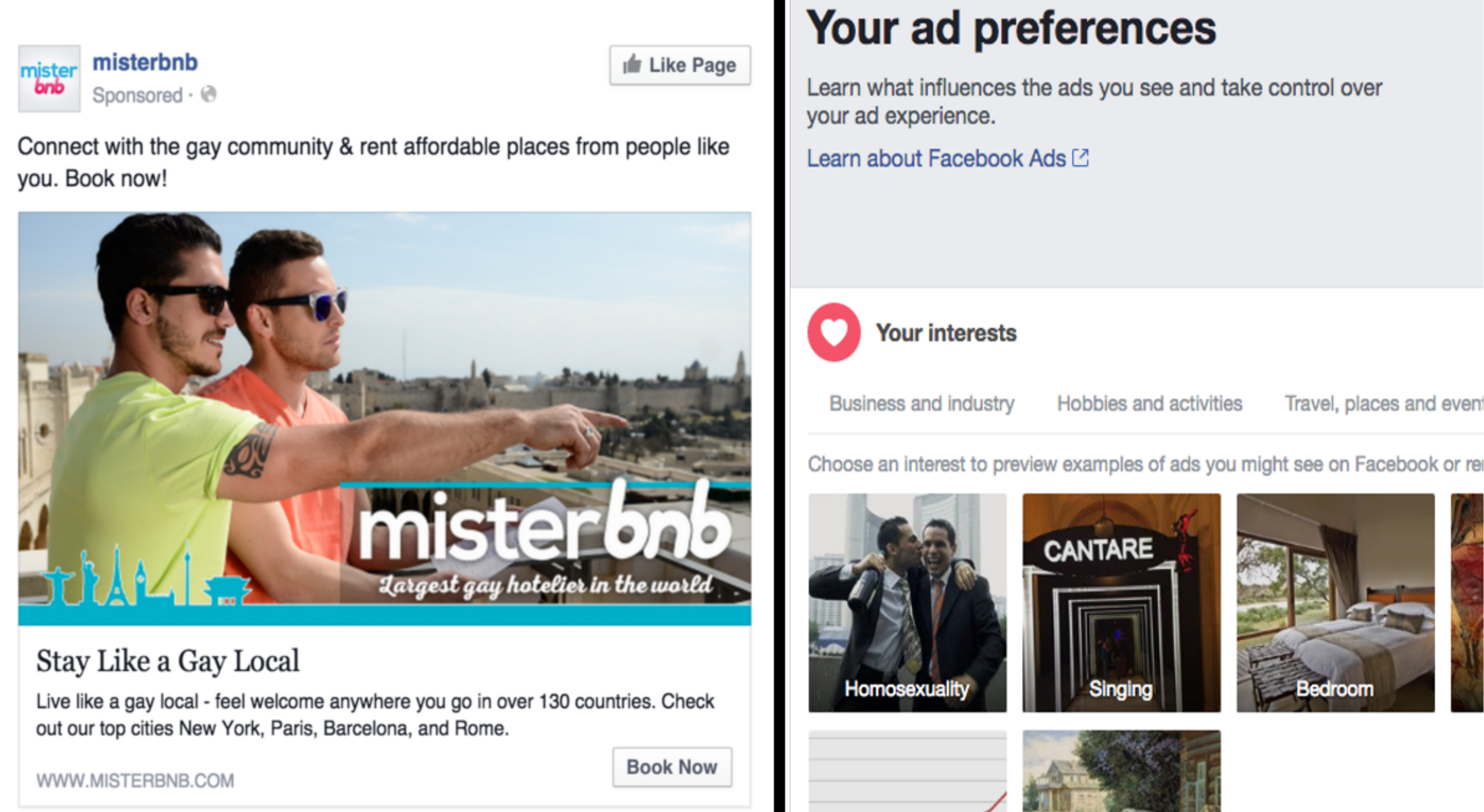}
\vspace{-1em}
\caption{Snapshot of an ad received by one of the authors of this paper \& ad preference list showing that FB inferred this person was interested in \textit{Homosexuality}.\vspace{-2em}}
\label{fig:ad1}
\end{figure}

FB labels users with the so-called ad preferences, which represent potential interests of users. FB assigns users different ad preferences based on their online activity within this social network. Advertisers running ad campaigns can target groups of users that have been assigned a particular ad preference (\textit{e.g.}, target FB users interested in \textit{``Starbucks''}). Some of these ad preferences suggest political opinions, sexual orientation, personal health, and other potentially sensitive attributes. In fact, an author of this paper received the ad shown in Figure \ref{fig:ad1} (left side). The text in the ad clearly reflects that the ad was targeting homosexual people. %The referenced} ad stated: \textit{``Connect with the gay community \& rent affordable places from people like you. Book Now''}. The company running the ad campaign is an online service targeting the gay community. 
The author had not explicitly defined his sexual orientation, but he discovered that FB had assigned him the \textit{``Homosexuality''} ad preference (see Figure \ref{fig:ad1} right side). The dataset collected for this research suggests that similar assignment of potentially sensitive ad preferences occurs much more broadly. For example, some landing pages associated with ads stored in our dataset include: \textit{iboesterreich.at} (political), \textit{gaydominante.com} (sexuality), \textit{elpartoestuyo.com} (health).

%This research focuses in FB due to an episode that happened to one of the authors of this paper. He received the ad shown on the left side of Figure \ref{fig:ad1}. The referred ad stated: \textit{``Connect with the gay community \& rent affordable places from people like you. Book Now''}. The company running the ad campaign is an online service targeting the gay community. 
%FB users are labeled with the so-called \emph{ad preferences}, which represent potential interests of the users. FB assigns a user different ad preferences based on her online activity within this social network and in third-party websites tracked by FB.  Advertisers running ad campaigns target groups of users that have been assigned a particular ad preference (e.g., target FB users interested in \textit{``Starbucks''}). The referred author found that \textit{``Homosexuality''} was one of the ad preferences FB had assigned him (see Figure \ref{fig:ad1} right side). He had neither explicitly defined his sexual orientation in any FB setting, nor granted explicit permission to FB to be targeted based on his sexual orientation. 
This episode illustrates that FB may be actually processing sensitive personal information, which is now prohibited under the EU GDPR without explicit consent, but it was neither allowed under some EU national data protection regulations prior to the GDPR. In May 2017, the French Data Protection Agency (DPA) fined Facebook with \euro150K arguing (among other things) that FB \textit{"collects sensitive data of the users without obtaining their explicit consent"}.\footnote{https://www.cnil.fr/en/facebook-sanctioned-several-breaches-french-data-protection-act} Similarly, In September 2017, the Spanish DPA fined FB \euro1.2M arguing (among other things) that FB \textit{``collects, stores and uses data, including specially protected data, for advertising purposes without obtaining consent''}.
\footnote{\url{https://techcrunch.com/2017/09/11/facebook-fined-e1-2m-for-privacy-violations-in-spain/}}
%\footnote{\url{http://www.agpd.es/portalwebAGPD/revista_prensa/revista_prensa/2017/notas_prensa/news/2017_09_11-iden-idphp.php}} %In both cases FB was fined based on national protection regulations prior to the GDPR.   

Motivated by all these events, this paper examines Facebook’s use of potentially sensitive data across 197 different countries in February 2019. The main goal of this paper is \textit{quantifying the portion of FB users that may have been assigned ad preferences linked to potentially sensitive personal data}. In addition, for the particular case of the 28 countries forming the EU, we analyze whether there has been some relevant reduction in the portion of users labelled with potentially sensitive ad preferences comparing three datasets collected in January 2018, October 2018 and February 2019 (5 months  before, 5 months after and 9 months after the GDPR was enacted, respectively). We also illustrate privacy and ethics risks that may be derived from the exploitation of sensitive FB ad preferences. Finally, we present a technical solution that allows users to remove in a simple way the potentially sensitive interests FB has assigned them. %We also study whether the potentially sensitive ads preferences assigned to users are frequently targeted by advertisers or instead they are very rarely used. To this end we compute which portion of the potentially sensitive ad preferences are actually semantically linked to the ads they receive.  

\begin{figure*}[!htb]
\centering
\includegraphics[width=.9\hsize]{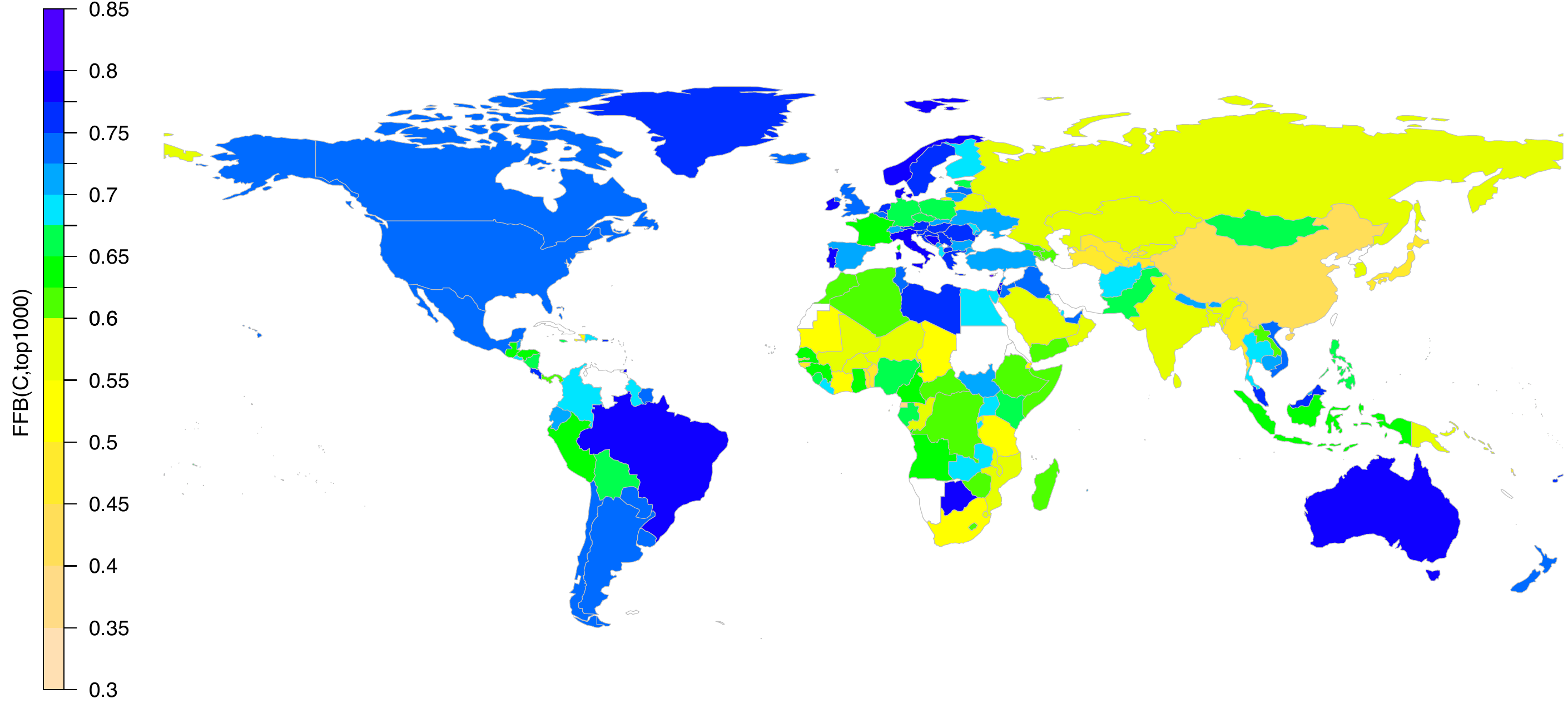}
%\vspace{-1em}
\caption{Choropleth map of the number of FB users assigned potentially sensitive ad preferences (FFB(C,1000)) for the 197 countries analyzed in the paper.\vspace{-1em}}
\label{fig:FFB_map}
\end{figure*}
\section{Background}

Advertisers configure their ads campaigns through the FB Ads Manager.\footnote{\url{https://www.facebook.com/ads/manager}} It allows advertisers to define the audience (\textit{i.e.}, user profile) they want to target with their advertising campaigns. It can be accessed through either a dashboard or an API. The FB Ads Manager offers advertisers a wide range of configuration parameters such as (but not limited to): \emph{location} (country, region, etc.), \emph{demographic parameters} (gender, age, etc.), \emph{behaviors} (mobile device, OS and/or web browser used, etc.), and \emph{interests} (sports, food, etc.). The \emph{interest} parameter is the most relevant for our work. It includes hundreds of thousands of possibilities capturing users' interest of any type. The FB Ads Manager provides detailed information about the configured audience. The most relevant element for our paper is the \emph{Potential Reach} that reports the number of monthly active users in FB matching the defined audience. 

In parallel, FB assigns to each user a set of ad preferences, \textit{i.e.}, a set of interests, derived from the data and activity of the user on FB. These ad preferences are indeed the interests offered to advertisers in the FB Ads Manager.\footnote{Given that interests and ad preferences refer to the same thing, we use these two terms interchangeably in the rest of the paper.} Therefore, if a user is assigned \textit{``Watches''} within her list of ad preferences, she will be a potential target of any FB advertising campaign configured to reach users interested in watches.

The dataset used in this work is obtained from the data collected with our FDVT web browser extension \cite{FDVT}. The FDVT main functionality is to inform FB users of the revenue they generate out of the ads they receive in FB. The FDVT collects (among other data) the ad preferences FB assigns to the user. It is important to note that FDVT users granted us explicit permission to use the collected information (in an anonymous manner) for research purposes. 

Finally, for any ad preference, we are able to query the FB Ads Manager API to retrieve the \textit{Potential Reach} (\textit{i.e.}, FB active users) associated to any FB audience. Hence, we are able to obtain the number of FB users in any country (or group of countries) that have been assigned a particular interest (or group of interests).

\section{Data and Methodology}
\label{sec:methodology}

We seek to quantify the number of FB users that have been assigned potentially sensitive ad preferences across 197 countries in February 2019. To this end we follow a two step process. 

First, we identify likely sensitive ad preferences within five of the relevant categories listed as \textit{Sensitive Personal Data} by the GDPR: racial or ethnic origin, political opinions, religious or philosophical beliefs, health, and sexual orientation. This paper reuses the list of 2092 potentially sensitive ad preferences we obtained in \cite{Jose_usenix} out of analyzing more than 126K unique ad preferences assigned 5.5M times to more than 4.5K FDVT users. 

To extract that list we first implemented an automatic process to reduce the list of 126K ad preferences to 4452 likely sensitive ad preferences. Next, a group of 12 panelists manually classified the 4452 ad preferences into sensitive, in case they could be assigned to some of the five sensitive categories referred above, or non-sensitive. Each ad preference received 5 votes, and we used majority voting \cite{narasimhamurthy2005theoretical} to classify each ad preference either as sensitive or non-sensitive. Overall, 2092 out of the 4452 ad preferences were labeled as sensitive. The complete list of the ad preferences classified as sensitive can be accessed via the FDVT site \footnote{\url{https://fdvt.org/usenix2018/panelists.html}}. We referred to this subset of 2092 ad preferences as the \textit{suspected sensitive subset}.  We collected this set in January 2018, and checked that 2067 out of these 2092 potentially sensitive ad preferences were still available within the FB Ads Manager in February 2019. 

Second, we leveraged the FB Ads Manager API to retrieve the portion of FB users in each country that had been assigned at least one of the Top N (with N ranging between 1 and 2067) potentially sensitive ad preferences from the suspected sensitive subset. In particular, we retrieve how many users in a given country are interested in \textit{ad preference 1} OR \textit{ad preference 2} OR \textit{ad preference 3}... OR \textit{ad preference N}. An example of this for N = 3 could be  \textit{``how many people in France are interested in Pregnancy OR Homosexuality OR Veganism''}. We have defined the following metric that we use in the rest of the paper  

-\emph{\textbf{FFB(C,N)}}: Percentage of FB users in country C that have been assigned at least one of the top N potentially sensitive ad preferences from the suspected sensitive subset. We note C may also refer to all the countries forming a particular region (e.g, EU, Asia-Pacific, America, etc.). FFB(C,N) is computed as the ratio between the number of FB users that have been assigned at least one of the top N potentially sensitive ad preferences and the total number of FB users in country C. Finally, it is important to note that the FB Ads Manager API only allows creating audiences with at most N = 1000 interests. Therefore, in practice, the maximum value of N we can use to compute FFB is 1000.

%-\emph{\textbf{FC(C,N)}}: This is the percentage of citizens in country C (or a region) that have been assigned at least one of the top N potentially sensitive ad preferences. It is computed as the ratio between the number of citizens that have been assigned at least one of the top N potentially sensitive ad preferences and the total population of country C. We use World Bank data to obtain countries' populations\footnote{\url{https://data.worldbank.org}}.

%The criterion to select the top N ad preferences out of the 2067 potentially sensitive ad preferences identified is popularity. Finally, it is important to note that the FB Ads Manager API only allows creating audiences with at most N = 1000 interests. Therefore, in practice, the maximum value of N we can use to compute FFB is 1000.

\section{Exposure of FB users to potentially sensitive ad preferences}
\label{sec:results}

We have computed the portion of FB users that have been assigned some of the 2067 potentially sensitive ad preferences within 197 different countries. Figure \ref{fig:FFB_map} shows a choropleth map of FFB(C,1000) for those countries in February 2019. 

If we consider the 197 all together, 67\% of FB users are tagged with some potentially sensitive ad preference. This portion of users actually corresponds to 22\% of citizens across the 197 analyzed countries according to the population data reported by the World Bank\footnote{\url{https://data.worldbank.org}}. However, FFB shows an important variation across countries. 

We find that the most impacted country is Malta where 82\% of FB users are assigned some potentially sensitive ad preference. Contrary, the least impacted country is Equatorial Guinea where 37\% of FB users are assigned potentially sensitive ad preferences. %Hence, we can conclude that in all the countries analyzed at least 37\% of FB users are assigned potentially sensitive ad preferences. %When looking at the FC results we find that the most and least impacted countries are United Arab Emirates and Turkmenistan with F(C,1000) equal to 72\%(Qatar 79\% February) and 0.35\%(0.42\% Turkmenistan February), respectively. 

\begin{table}[t]
\centering
%\small
\begin{adjustbox}{width=\columnwidth}
\begin{tabular}{|l|cc|}
\hline
    indicator & correlation FFB & p\_value \\
\hline
    FB penetration & 0.544 & 2.2e-16 \\
    Expected Years of School & 0.444 & 7.249e-09 \\
    \begin{tabular}{@{}l@{}}Access to a mobile phone or \\ internet at home (\% age 15+) \end{tabular}
     & 0.395  & 1.478e-06 \\
    GDP per capita (current USD) & 0.381 & 5.733e-08 \\
    Voice and Accountability & 0.372  & 1.142e-07 \\
    %Net migration & 0.184 & 0.013 \\
    Birth rate, crude (per 1,000 people) & -0.455 & 4.922e-11 \\
    \hline
\end{tabular}
\end{adjustbox}
\caption{Pearson correlation and p\_value between FFB and six socioeconomic development indicators of the country.\vspace{-3em}}
\label{table:correlations}
\end{table}

More interesting, an overview of the map seems to suggest that western countries have a higher exposure to potentially sensitive ad preferences compared to Asian and African countries.  To quantify these effects we have computed the Pearson correlation of the FFB metric with the following socio-economic indicators: (i) FB penetration, (ii) expected years of school; %i.e., \textit{the sum of age-specific enrollment rates between ages 4 and 17}; 
(iii) access to a mobile phone or internet at home; (iv) GDP per capita; (v) voice and accountability;%, i.e, \textit{it captures perceptions of the extent to which a country's citizens are able to participate in selecting their government, as well as freedom of expression, freedom of association, and a free media}  %(vi) net migration 
and (vi) birth rate. Note that Western developed countries shows higher values in all the indicators but birth rate. Hence our hypothesis is that we will find positive correlation between FFB and all the indicators but birth rate. %for which we expected to find a significant negative correlation.  
Table \ref{table:correlations} shows the results of the referred correlations. Note that in all the cases the results are statistically significant since the highest p-value is 1.478e-06. 

The results Table \ref{table:correlations} corroborate our hypothesis since all the indicators but birth rate are positively correlated with FFB. %We find strong correlations with all the indicators except net migration. 
In summary, the results validate our initial observation that FB users in western developed countries are more exposed to be labelled with sensitive ad preferences than users in Africa and Asia. It is interesting to observe that in the case of South-America we observe a similar pattern in which the most powerful economies and developed countries such as Brazil, Chile and Argentina shows higher exposure to sensitive ad preferences than other countries in South-America.

\section{Exposure of FB users to very sensitive ad preferences}
Although legislation tries to define what sensitive data is, some people might think that not all different sensitive data items are equally sensitive. For instance, data revealing sexual orientation from somebody could be considered more sensitive than, for example, data showing that one user may be affected by a flu. Therefore, the level of sensitivity of our list of interests could vary depending on the importance given by someone. 

In this section, we zoom in our analysis in a narrowed list of interests that match undoubtedly with the definition of the GDPR for the case of sensitive personal data. We examined a subset of 15 ad preferences that have been verified by an expert from the Spanish DPA as initially not compliant with the GDPR definition of sensitive personal data.

We retrieve the portion of users in FB for each of the 15 expert-verified ad preferences and the aggregation of them. Since it is unfeasible to show the results for each of the countries within the paper, we have grouped them into five continents: Africa, America, Asia, Europe and Oceania. To obtain the desegregated results for each of the 197 countries we refer the reader to the following external link.\footnote{\url{https://fdvt.org/world_sensitivities_2019/display_sensitivities.html}}

\begin{table}[t]
\centering
\resizebox{\hsize}{!}{
\begin{tabular}{|l|c|c|c|c|c|c|}
  \hline
  ad preference & Africa & America & Asia & Europe & Oceania & World \\
  \hline
  ALTERNATIVE MEDICINE & 3.40 & 11.35 & 3.27 & 7.17 & 10.82 & 6.26 \\ 
  BIBLE & 13.28 & 14.65 & 6.31 & 8.13 & 14.61 & 9.68 \\ 
  BUDDHISM & 2.87 & 5.38 & 10.36 & 4.13 & 7.19 & 7.23 \\ 
  FEMINISM & 3.22 & 9.27 & 2.08 & 6.52 & 10.84 & 5.01 \\ 
  GENDER IDENTITY & 0.08 & 0.46 & 0.07 & 0.20 & 0.60 & 0.21 \\ 
  HOMOSEXUALITY & 2.66 & 7.93 & 2.27 & 6.07 & 8.48 &  4.57 \\ 
  ILLEGAL IMMIGRATION & 0.26 & 0.15 & 0.02 & 0.03 & 0.07 &  0.08 \\ 
  JUDAISM & 11.06 & 3.72 & 1.91 & 2.24 & 2.44 &  3.33 \\ 
  LGBT COMMUNITY & 3.93 & 13.89 & 5.39 & 11.94 & 14.82 & 8.79 \\ 
  NATIONALISM & 1.82 & 1.11 & 1.28 & 1.32 & 0.95 & 1.28 \\ 
  ONCOLOGY & 1.30 & 1.33 & 0.38 & 0.84 & 0.97 & 0.81 \\ 
  PREGNANCY & 11.75 & 19.17 & 11.58 & 17.09 & 21.41 & 14.71 \\ 
  REPRODUCTIVE HEALTH & 0.36 & 0.24 & 0.17 & 0.07 & 0.09 & 0.19 \\ 
  SUICIDE PREVENTION & 0.05 & 0.30 & 0.03 & 0.08 & 1.02 &  0.13 \\
  VEGANISM & 5.97 & 14.18 & 6.83 & 16.98 & 22.78 & 10.61 \\ 
  \hline
  UNION & 30.43 & 40.66 & 27.62 & 38.25 & 46.92 &  33.45 \\ 
  \hline
\end{tabular}}
\caption{Percentage of FB users (FFB) within Africa, America, Asia, Europe and Oceania assigned some sensitive ad preferences from a list of 15 expert-verified sensitive ad preferences as non-GDPR compliant. Last column "World" shows FFB for the aggregation of all 197 considered countries. Last row shows the result for the 15 ad preferences aggregated. \vspace{-3em}}
\label{tab:very_sensitive}
\end{table}

Table \ref{tab:very_sensitive} shows FFB for each of the expert-verified sensitive ad preferences within the five continents. In addition, the last row referred to as \textit{Union} shows the aggregated results considering all the 15 interests within a group, while the last column \textit{World} depicts the overall results considering all 197 countries. The results shows that when considering all the 197 countries 33\% of FB users, which corresponds to almost 11\% of citizens within those countries, have been labeled with some of the 15 sensitive interests in the table. As it was expected from the correlation results depicted in the previous section, Asia and Africa are showing the lowest values of FFB (27.62\% and 30.43\%, respectively). The exposition of FB users grows up to 38.25\% , 40.66\% and 46.92\%  for Europe, America and Oceania, respectively. %For the case of Europe and America, when we limit the scope of the analysis to either EU28 or US\&CA the FFB value grows by 1.5 and 5 percentage points, respectively.

If we look in detail some of the ad preferences in the table, we observe that the portion of users across the 197 countries labeled with the ad preference homosexuality is almost 5\%. This number doubles for the ad preference bible (intimate related to one particular religious belief), and grows up to almost 15\% for pregnancy. 

\section{Comparison of EU FB users exposure to potentially sensitive ad preferences before and after GDPR enforcement}

This section aims to analyze whether the GDPR enforcement had some effect on the utilization of potentially sensitive ad preferences to label FB users in the EU. To that end we compare the exposure of EU users to potentially sensitive ad preferences in January 2018 \cite{Jose_usenix} (5 months before the GDPR was enforced) to the exposure measured in October 2018 and February 2019 (5 and 9 months after the GDPR was enforced, respectively). 

The first relevant change is that Facebook had removed 19 ad preferences in October 2018 and 25 in February 2019 from the set of 2092 potentially sensitive ad preferences we retrieved in January 2018. Although this is a negligible amount, it is worth noting that five of the removed ad preferences are: Communism, Islam, Quran, Socialism and Christianity. These five ad preferences were included in an initial set of 20 ad preferences verified by the DPA expert as very sensitive. Hence, it seems FB is starting to consider some very sensitive interests as too invasive and has decided to remove them from its advertising platform.

\begin{figure}[t]
    \centering
    \includegraphics[width=\columnwidth]{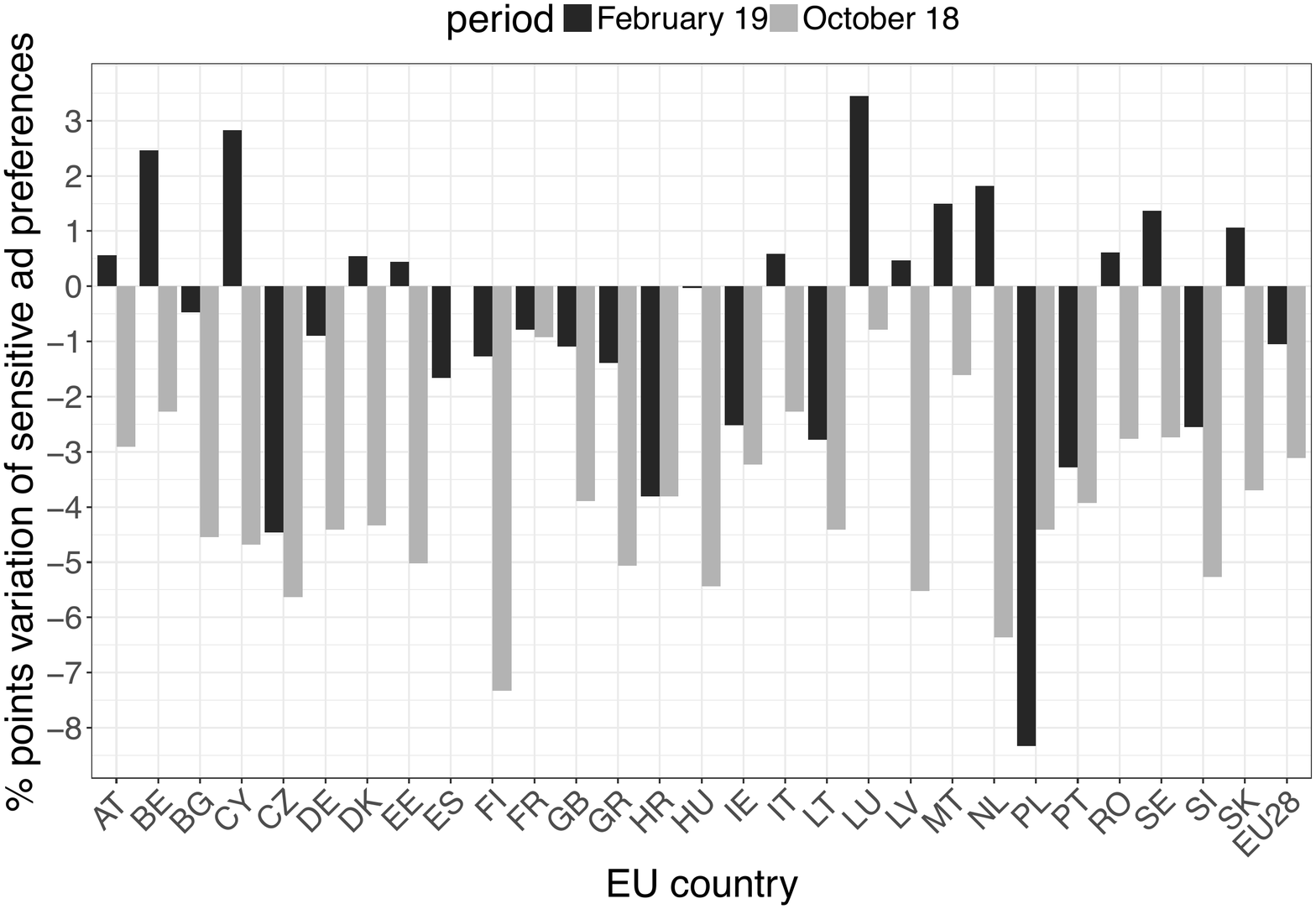}
    \vspace{-2em}
    \caption{Variation of FFB in percentage points for each EU country between: (i) the data obtained in January 2018 and October 2018 (5 months before and 5 months after the GDPR was enacted) represented by the grey bar; (ii) the data obtained in January 2018 and February 2019 (5 months before and 9 months after the GDPR was enacted) represented by the black bar. The last label (EU28) represents the results for all EU countries together.\vspace{-2em}}
    \label{fig:variation}
\end{figure}

Figure \ref{fig:variation} shows the FFB difference in percentage points between the results obtained in January 2018 and October 2018 (grey bar); and between January 2018 and February 2019 (black bar) across the 28 EU countries, and the EU aggregated labeled as EU28.

If we consider the results of October 2018, we observe that the portion of users labelled with potentially sensitive ad preferences was lower in all EU countries but Spain after the GDPR enforcement (\textit{i.e.,} compared to the data obtained in January 2018). However, the aggregated EU reduction is rather small, only 3 percentage points. The largest reduction is 7.33 percentage points in the case of Finland.

The slight GDPR effect observed in the results obtained in October 2018 seems to disappear when we observe the results from February 2019. There are 13 countries where the portion of users labelled with potentially sensitive data is higher in February 2019 as compared to January 2018. Overall, the aggregated results shows that the portion of users labeled with potentially sensitive ad preferences in February 2019 is only 1\% less than in January 2018.

In summary, FB seems to have adopted some steps to eliminate few very invasive ad preferences, but the overall impact of the GPDR to prevent FB of using potentially sensitive ad preferences for advertising purposes is negligible.

\section{Ethics and privacy risks associated with sensitive personal data exploitation}
\label{sec:discussion}

\begin{table}[t]
\centering
%\small
\begin{adjustbox}{width=\columnwidth}
\begin{tabular}{|llc|llc|}
  \hline
code & country & homosexuality & code & country & homosexuality\\ 
  \hline
 AF & AFGHANISTAN & 12.31 & BN & BRUNEI & 5.24\\
 MR & MAURITANIA & 0.99& NG & NIGERIA & 2.35\\
 QA & QATAR & 2.35 & SA & SAUDI ARABIA & 2.08\\
 SO & SOMALIA & 1.44 & YE & YEMEN & 1.08\\
 PK & PAKISTAN & 1.54 & IQ & IRAQ & 3.20\\
 AE & UNITED ARAB EMIRATES & 3.00 & & &\\
 
 \hline
\end{tabular}
\end{adjustbox}
\caption{Percentage of FB users (FFB) tagged with the interest \emph{Homosexuality} in countries where being homosexual may lead to death penalty. Note we do not include Iran and Sudan since FB is not providing information for those countries.\vspace{-3em}}
\label{table:homosexuality_penal}
\end{table}

The possibility of reaching users labeled with potentially sensitive personal data enables the use of FB ads campaigns to attack (\textit{e.g.,} hate speech) specific groups of people based on sensitive personal data (ethnicity, sexual orientation, religious beliefs, etc.). Even worse, in \cite{Jose_usenix}, we performed a ball-park estimation showing that in average an attacker could retrieve personal identifiable information (PII) of users tagged with some sensitive ad preference through a phishing-like attack \cite{phising_survey} at a cheap cost ranging between \euro0.015 and \euro1.5 per user, depending on the success ratio of the attack. Following, we describe other potential risks associated to sensitive ad preferences.

Recently, a journalist of the Washington Post wrote an article to denounce her own experience after she become pregnant.\footnote{\url{https://www.washingtonpost.com/lifestyle/2018/12/12/dear-tech-companies-i-dont-want-see-pregnancy-ads-after-my-child-was-stillborn/}} It seems FB algorithms inferred that situation out of some actions she performed while browsing in Facebook. Probably FB labelled her with the ad preference \textit{"pregnancy"} or some other similar and she started to receive pregnancy-related ads. Unfortunately, the journalist had a stillbirth but she kept receiving ads related to pregnancy, which exposed here to a very uncomfortable experience.

Another serious risk, which in our opinion is extremely worrying, is linked to the fact that many FB users are tagged with the interest \textit{"Homosexuality"} in countries where being homosexual is illegal and may even be punished with the death penalty. There are still 78 countries in the world where the homosexuality is penalized\footnote{\url{https://ilga.org/downloads/2017/ILGA_WorldMap_ENGLISH_Criminalisation_2017.pdf}} and few of them such where Death Penalty is the maximum punishment. Table \ref{table:homosexuality_penal} shows the FFB metric results only considering the interest \textit{"Homosexuality"} in countries that penalize homosexuality with the death penalty. For instance, in the case of Saudi Arabia we found that FB assigns the ad preference \textit{"Homosexuality"} to 540K people (2.08\% of FB users in that country). In the case of Nigeria 620K (2.35\% of FB users in that country). 

We acknowledge the debate regarding what is sensitive and what is not is a complex one. However, we believe FB should take immediate actions to avoid worrying and painful situations like the one exposed in this section, in which FB may unintentionally expose users to serious risks. For instance, a straightforward action should be stop using the ad preference \textit{"Homosexuality"} (or similar ones) in countries where being homosexual is illegal.

\section{FDVT extension to allow users removing potentially sensitive ad preferences}
\label{sec:fdvt_extension}
% Our current research line is in the field of web transparency to create awareness among Internet users regarding the way online services exploit their personal data in the area of online adverting. The current research could be used to extend the FDVT with a new functionality that let users know that FB is exploiting sensitive personal data for advertising purposes.  To this end, our goal is to provide FDVT users with the list of sensitive ad preferences that FB has assigned them over the time. 

%\begin{figure}
%\centering
%	\includegraphics[width=.75\columnwidth]{./figures/boxplot_performance_5k.eps}\hfill
%	\caption{AUC, precision, recall and F-score for the optimal threshold to automatically classify an ad preference as sensitive or non-sensitive. The figures shows the results obtained from 5000 iterations across different randomly chosen training and validation data subsets.}
%	\label{fig:test_training}
%\end{figure} 

The results reported in previous sections motivate a need for solutions that make users aware of the use of sensitive personal data for advertising purposes. In addition, it is also important to empower them to remove in a very simple manner those sensitive ad preferences they do not fill comfortable with. Unfortunately, the existing process FB offers is unknown and complex for most users.%\footnote{\url{https://www.facebook.com/ads/preferences/}} 
To this end, we have extended the FDVT browser extension to: (i) inform users about the potentially sensitive ad preferences that FB has assigned them, both the active ones but also those ones assigned in the past that are not currently active; (ii) allow users to remove with a single click either all the active sensitive ad preferences or those individual ones users do not fill comfortable with. 

\begin{figure}[t]
    \centering
    \includegraphics[width=\columnwidth]{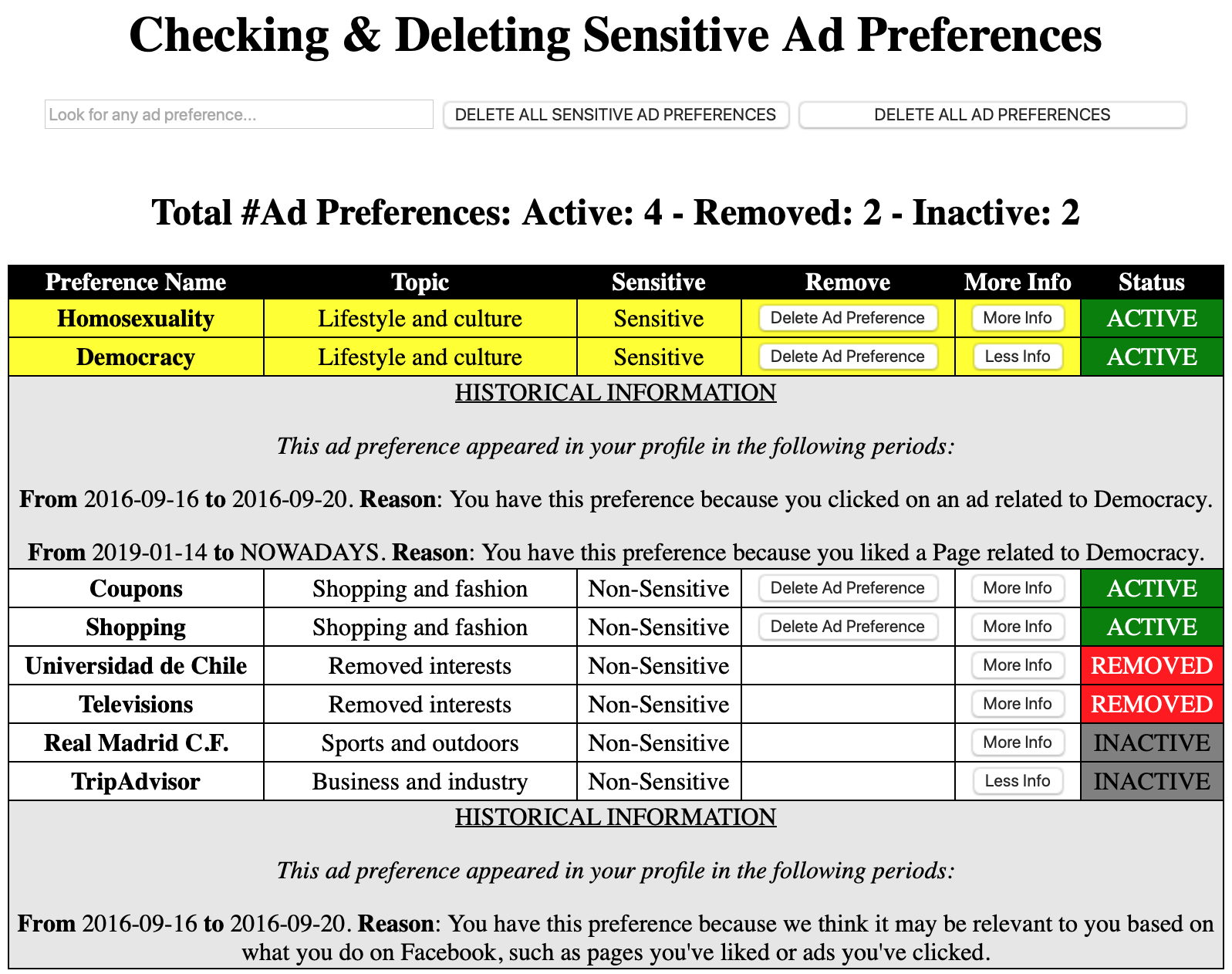}
    \vspace{-2em}
    \caption{Snapshot of FDVT new feature to allow users deleting sensitive ad preferences.\vspace{-2em}}
    \label{fig:sens_page}
\end{figure}

We have introduced a new button in the FDVT extension interface with the label \textit{''Sensitive FB Preferences''}. When a user clicks on that button, we display a page listing at the top the potentially sensitive ad preferences included in the user's ad preference set (both the active ones and inactive ones). Figure \ref{fig:sens_page} shows an example of this page. We provide the following information for each ad preference: $(i)$ Ad preference name, $(ii)$ Topic and, $(iii)$ Sensitive, whether the ad preference is potentially sensitive (highlighted in yellow) or not. In addition, next to each ad preference there is a button \textit{Delete Ad Preference} to individually remove those ad preferences. Moreover, we provide another button \textit{More Info} to individually display the historical information for the ad preference, which includes the period(s) when the ad preference has been active and the reason why FB has assigned that ad preference to the user. Finally, at the top of the page we include a search bar to look for specific preferences and two buttons: \textit{Delete All Sensitive Ad Preferences} and \textit{Delete All Ad Preferences} to remove all currently active potentially sensitive ad preferences and all currently active, respectively.%\vspace{-1.5em}

\section{Related work}

 We published a prior paper \cite{Jose_usenix} in which we already analyzed the use of sensitive information on Facebook. That paper just focuses on the European Union few months before the GDPR was enacted. The research community asked us in various forums that it would be interesting to further extend our analysis to: (i) cover the use of sensitive information in Facebook worldwide and not just in the EU, and (ii) understand the potential impact that the GDPR could have on reducing the exposure of users to sensitive ad preferences. This paper covers both requests and, in addition, it adds two more contributions: (i) we present two clear scenarios in which the use of sensitive ad preferences could have serious consequences for the users;  and (ii) we introduce an improvement of the FDVT that allows users to remove in a simple way potentially sensitive ad preferences they do not like. Hence, this paper notably extends our previous work.

There are also few previous works in the literature that address issues associated with sensitive personal data in online advertising, as well as some recent works that analyze privacy and discrimination issues related to FB advertising and ad preferences.% While is a large corpus of literature addressing privacy issues associated to online advertising and identification of users based on personal data \cite{de2013unique}\cite{de2015unique},

Carrascosa et al. \cite{Carrascosa} propose a new methodology to quantify the portion of targeted ads received by Internet users while they browse the web. They create bots, referred to as \textit{personas}, with very specific interest profiles (e.g., persona interested in cars) and measure how many of the received ads actually match the specific interest of the analyzed persona. They create personas based on sensitive personal data (\textit{e.g.,} health) and demonstrate that they are also targeted with ads related to the sensitive information used to create the persona's profile.  

Castellucia et al. \cite{castelluccia2012betrayed} show that an attacker that gets access (\textit{e.g.,} through a public WiFi network) to the Google ads received by a user could create an interest profile that could reveal up to 58\% of the actual interests of the user. The authors state that if some of the unveiled interests are sensitive, it could imply serious privacy risks for users. %In the case of FB, an attacker does not need to retrieve any interest from the user, but using Facebook ad preferences to reach the group of users she is targeting in the identification attack.

Venkatadri et al. \cite{EURECOM+5420} and Speicher et al. \cite{speicherpotential} exposed privacy and discrimination vulnerabilities related to FB advertising. In \cite{EURECOM+5420}, the authors demonstrate how an attacker can use Facebook third-party tracking JavaScript to retrieve personal data (\textit{e.g.,} mobile phone numbers) associated with users visiting the attacker's website. Moreover, in \cite{speicherpotential} they demonstrate that sensitive FB ad preferences can be used to apply negative discrimination in advertising campaigns (\textit{e.g.,} excluding people based on their race). The authors also show that some ad preferences that initially may not seem sensitive could be actually used to discriminate in advertising campaigns (\textit{e.g.,} excluding people interested in \textit{Blacknews.com} that are potentially black people). %Aligned to this finding our algorithm to automatically identify ad preferences uses a semantic similarity approach that will detect as sensitive many ad preferences that initially would not be labeled as such. 

%Andreou et al. \cite{andreou2018investigating} analyze whether the reasons FB uses to explain why a user is targeted with an ad are aligned with the actual audience the advertiser is targeting. To do this, they analyze the explanation that Facebook includes in each delivered ad referred to as \textit{``Why Am I Seeing this Ad''}. This explanation describes the target audience associated with the delivered ad.  Out of the analysis of 79 ads, they conclude that in many cases the provided explanations are incomplete and sometimes misleading. They also perform a qualitative analysis related to the ad preferences assigned to FB users based on a small dataset including 9K ad preferences distributed across 35 users. They conclude that the reasons why ad preferences are assigned are vague.

\section{Conclusion}

Facebook offers advertisers the option to commercially exploit potentially sensitive information to perform tailored ad campaigns. This practice lays, in the best case, within a gray legal area according to the recently enforced GDPR. Our results reveal that 67\% of FB users (22\% of citizens) worldwide are labeled with some potentially sensitive ad preference. Interestingly, users in rich developed countries present a significantly higher exposure to be assigned sensitive ad preferences. Our work also reveals that the enforcement of the GDPR had a negligible impact on FB regarding the use of sensitive ad preferences within the EU. We believe it is urgent that stakeholders within the online advertising ecosystem (\textit{i.e.,} advertisers, ad networks, publishers, policy makers, etc.) define an unambiguous list of personal data items that should not be used anymore to protect users from potential privacy risks as those ones described in this paper.
%\vspace{-1em}

\section*{Acknowledgment}
This work was partially funded by: (i) the Ministerio de Economía, Industria y Competitividad, Spain, and the European Social Fund (EU), under the Ramón y Cajal programme (grant RyC-2015-17732), and the Project TEXEO (Grant TEC2016-80339-R), (ii)  the European H2020 Project SMOOTH (Grant 786741), (iii) the Ministerio de Educación, Cultura y Deporte, Spain, through the FPU programme (Grant FPU16/05852), and (iv) the Community of Madrid synergic project EMPATIA-CM (Grant Y2018/TCS-5046).

%\balance
\bibliographystyle{ACM-Reference-Format}
\bibliography{references}

% that's all folks
\end{document}